\documentclass[preprint,onecolumn,aps,pre,eqsecnum,longbibliography,showpacs,showkeys,a4paper]{revtex4-1}
\usepackage[latin9]{inputenc}
\usepackage{color}
\definecolor{note_fontcolor}{rgb}{0.800781, 0.800781, 0.800781}
\usepackage{units}
\usepackage{amsmath}
\usepackage{amssymb}
\usepackage{graphicx}
\usepackage[unicode=true,
 bookmarks=true,bookmarksnumbered=true,bookmarksopen=true,bookmarksopenlevel=1,
 breaklinks=true,pdfborder={0 0 0},backref=false,colorlinks=true]
 {hyperref}
\hypersetup{
 pdfauthor={AINS}}

\makeatletter

\newenvironment{lyxgreyedout}
  {\textcolor{note_fontcolor}\bgroup\ignorespaces}
  {\ignorespacesafterend\egroup}

\usepackage{graphicx}

\DeclareGraphicsExtensions{.png,.pdf,.eps}
\graphicspath{{.}}

\makeatother

\begin{document}

\title{Emergent quantum mechanics without wave functions}

\author{Johannes \surname{Mesa Pascasio}\textsuperscript{1,2}}

\email[E-mail: ]{ains@chello.at}

\homepage[Visit: ]{http://www.nonlinearstudies.at/}

\author{Siegfried \surname{Fussy}\textsuperscript{1}}

\email[E-mail: ]{ains@chello.at}

\homepage[Visit: ]{http://www.nonlinearstudies.at/}

\author{Herbert \surname{Schwabl}\textsuperscript{1}}

\email[E-mail: ]{ains@chello.at}

\homepage[Visit: ]{http://www.nonlinearstudies.at/}

\author{Gerhard \surname{Grössing}\textsuperscript{1}}

\email[E-mail: ]{ains@chello.at}

\homepage[Visit: ]{http://www.nonlinearstudies.at/}

\affiliation{\textsuperscript{1}Austrian Institute for Nonlinear Studies, Akademiehof\\
 Friedrichstr.~10, 1010 Vienna, Austria}

\affiliation{\textsuperscript{2}Atominstitut, TU Wien, Operng.~9, 1040 Vienna,
Austria}

\affiliation{\vspace*{2cm}
}

\date{\today}
\begin{abstract}
We present our model of an Emergent Quantum Mechanics which can be
characterized by \textquotedblleft realism without pre-determination\textquotedblleft .
This is illustrated by our analytic description and corresponding
computer simulations of Bohmian-like \textquotedblleft surreal\textquotedblleft{}
trajectories, which are obtained classically, i.e.~without the use
of any quantum mechanical tool such as wave functions. However, these
trajectories do not necessarily represent ontological paths of particles
but rather mappings of the probability density flux in a hydrodynamical
sense. Modelling emergent quantum mechanics in a high-low intesity
double slit scenario gives rise to the \char`\"{}quantum sweeper effect\char`\"{}
with a characteristic intensity pattern. This phenomenon should be
experimentally testable via weak measurement techniques.%
\begin{lyxgreyedout}
\global\long\def\VEC#1{\mathbf{#1}}

\global\long\def\d{\,\mathrm{d}}

\global\long\def\e{{\rm e}}

\global\long\def\meant#1{\left<#1\right>}

\global\long\def\meanx#1{\overline{#1}}

\global\long\def\mpbracket{\ensuremath{\genfrac{}{}{0pt}{1}{-}{\scriptstyle (\kern-1pt +\kern-1pt )}}}

\global\long\def\pmbracket{\ensuremath{\genfrac{}{}{0pt}{1}{+}{\scriptstyle (\kern-1pt -\kern-1pt )}}}

\global\long\def\p{\partial}
\end{lyxgreyedout}

\end{abstract}

\keywords{neutron interferometry, double slit}

\maketitle

\section{Beam attenuation}

We have studied beam attenuation experiments in neutron interferometry
as introduced by~\cite{Summhammer.1987stochastic} for extreme attenuation
values. We use the \emph{transmission factor} $a$ as the beam's transmission
probability~\cite{Groessing.2014attenuation}, in the cases of a
(deterministic) chopper wheel with $a=\nicefrac{t_{\mathrm{open}}}{t_{\mathrm{open}}+t_{\mathrm{closed}}}$
given by the temporal open-to-closed ratio, and a (stochastic) semitransparent
material with $a=\nicefrac{I}{I_{0}}$ being the relation of the intensity
$I$ with absorption compared to the intensity $I_{0}$ without. Thus
the beam modulation behind the interferometer is obtained in two ways.
With $\varphi$ denoting the phase difference, the intensities are
given for respective cases by
\begin{eqnarray}
I_{\mathrm{det}} & \propto & 1+a+2a\cos\varphi\label{eq:3}\\
I_{\mathrm{stoch}} & \propto & 1+a+2\sqrt{a}\cos\varphi\label{eq:4}
\end{eqnarray}
Although the same number of particles is observed in both cases, in
Eq.~\eqref{eq:3} the contrast of the interference pattern is proportional
to $a$, whereas in Eq.~\eqref{eq:4} it is proportional to $\sqrt{a}$.
In our calculations  we assume stochastic type attenuation.
\begin{figure}
\noindent \centering{}\includegraphics[width=1\columnwidth]{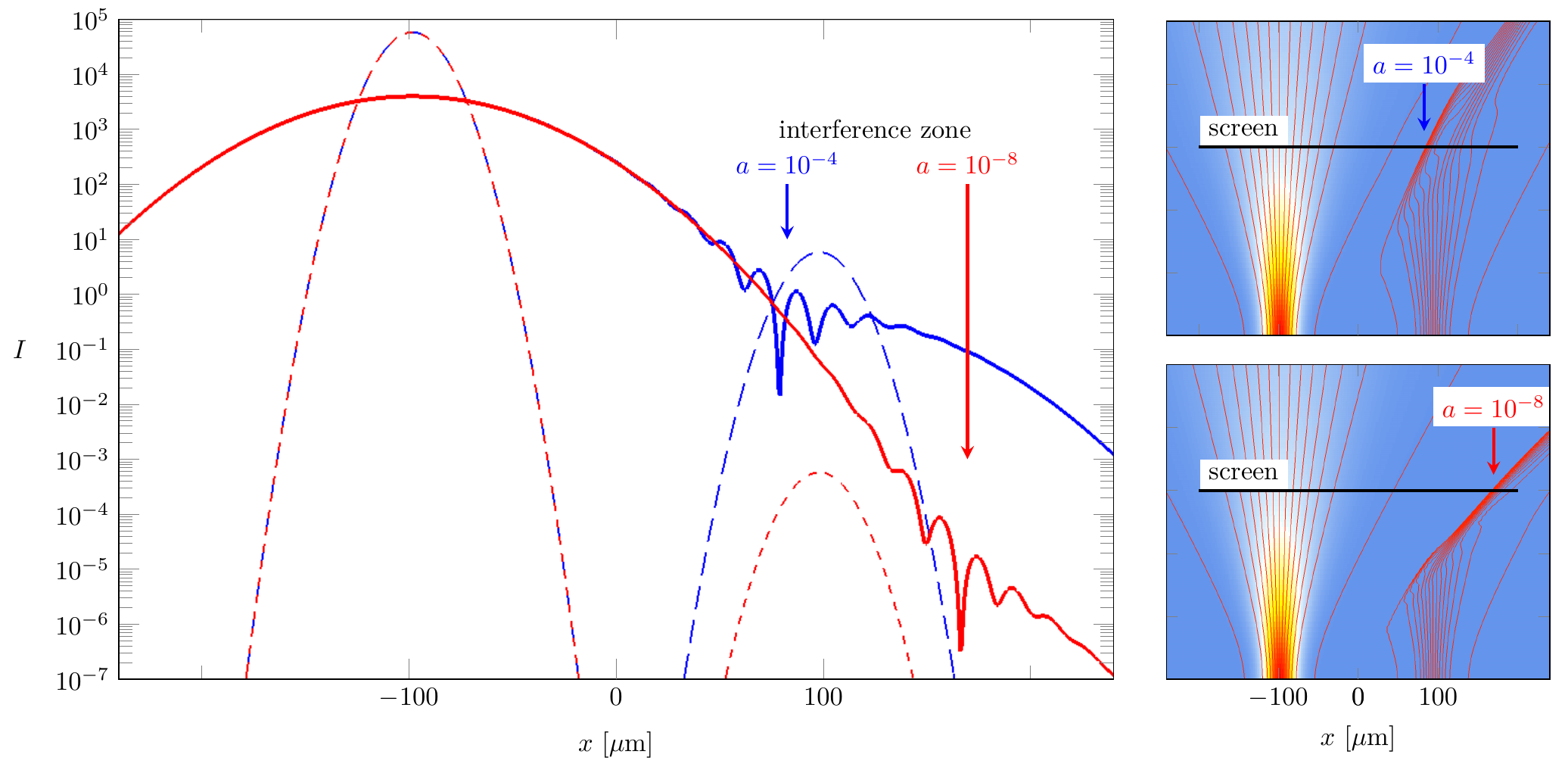}\caption{\textit{(Left)} Dashed initial distributions for the cases of $a=10^{-4}$
(blue) and $a=10^{-8}$ (red), respectively, develop into distributions
clearly showing interference phenomena which have been ``swept aside''
far to the right. This sweeper effect is due to the explicit appearance
of the nonlinear structure of the probability density current in these
domains for very low values of $a$. \textit{(Right)} Probability
density distributions $I$ emanating from the double slit with transmission
factor $a=10^{-4}$ (top) and $a=10^{-8}$ (bottom), respectively.\label{fig:1}}
\end{figure}

\section{The sweeper effect}

The probability density distribution for the right slit exhibits marked
signs of interference effects due to the compressed wave superpositions
within the bunching area caused by the sweeper effect. For transmission
factors below $a\lesssim10^{-4}$, new effects appear which are not
taken into account in linear extrapolation of expectations based on
higher-valued transmission factors.

With ever lower values of $a$, one can see a steadily growing tendency
for the attenuated beam to become swept aside. In our model, this
phenomenology is explained by processes of diffusion, due to the presence
of accumulated vacuum heat (i.e.\ kinetic energy) mainly in the ``strong''
beam. The sweeper effect is thus the result of the vacuum heat sweeping
aside the very low intensity beam, with a sharp boundary defined by
the balancing of the osmotic momenta coming from the two beams, respectively.
Due to the different $a$-dependencies in Eqs.~\eqref{eq:3} and~\eqref{eq:4},
we observe different shifts of the intensity patterns and correspondingly
different angles of the above-mentioned boundary.

As was already pointed out by Luis and Sanz~\cite{Luis.2015what},
the resulting trajectories such as those of Fig.~\ref{fig:1} can
be understood as a nonlinear effect that is not usually considered
in standard quantum mechanics, but explainable in the Bohmian picture.
There, it is the structure of the velocity field which is genuinely
nonlinear and therefore allows for the emergence of the type of trajectory
behavior observed. Moreover, in our approach, the emergence of the
trajectories of Fig.~\ref{fig:1} can be traced back to the non-vanishing
of the entangling current $\frac{\hbar}{m}\left(R_{1}\nabla R_{2}-R_{2}\nabla R_{1}\right)$~\cite{Groessing.2015dice}.
\begin{acknowledgments}
We thank Maurice de Gosson and Jan Walleczek for many enlightening
discussions and the Fetzer Franklin Fund for partial support of the
current work.
\end{acknowledgments}

\section*{References}

\bibliographystyle{utphys}
\bibliography{ains-reduced}

\end{document}